# Halogenated MXenes with Electrochemically Active Terminals for High Performance Zinc Ion Batteries

Mian Li[1, 2 #], XinLiang Li[3 #], Guifang Qin[1, 2, 5 #], Kan Luo[1, 2], Jun Lu[4], Youbing Li[1, 2], Guojin Liang[3], Zhaodong Huang[3], Lars Hultman[4], Per Eklund[4], Per O.Å. Persson[4], Shiyu Du[1,2], Zhifang Chai[1,2], Chunyi Zhi[3*], Qing Huang[1,2*]

1. Engineering Laboratory of Advanced Energy Materials, Ningbo Institute of Materials Technology& Engineering, Chinese Academy of Sciences, Ningbo, Zhejiang, 315201, China; 2. Qianwan Institute of CNiTECH，Zhongchuangyi Road, Hangzhou bay District, Ningbo, Zhejiang, 315336; 3. Department of Materials Science and Engineering, City University of Hong Kong, 83 Tat Chee Avenue, Kowloon, P. R. China; 4. Department of Physics, Chemistry, and Biology (IFM), Linköping University, 58183, Linköping, Sweden; 5. University of Chinese Academy of Sciences, 19A Yuquan Rd, Shijingshan District, Beijing, 10049, China

**Abstract**

The class of two-dimensional metal carbides and nitrides known as MXenes offer a distinct manner of property tailoring for a wide range of applications. The ability to tune the surface chemistry for expanding the property space of MXenes is thus an important topic, although experimental exploration of new surface terminals remains a challenge. Here, we synthesized $Ti_3C_2$ MXene with unitary, binary and ternary halogen terminals, e.g. –Cl, –Br, –I, –BrI and –ClBrI, to investigate the effect of surface chemistry on the properties of MXenes. The electrochemical activity of Br and I element result in the extraordinary electrochemical performance of the MXenes as cathodes for aqueous zinc ion batteries. The –Br and –I containing MXenes, e.g. $Ti_3C_2Br_2$ and $Ti_3C_2I_2$, exhibit distinct discharge platforms with considerable capacities of 97.6 mAh·g$^{-1}$ and 135 mAh·g$^{-1}$. $Ti_3C_2(BrI)$ and $Ti_3C_2(ClBrI)$ exhibit dual discharge platforms with capacities of 117.2 mAh·g$^{-1}$ and 106.7 mAh·g$^{-1}$. In contrast, the previously discovered MXenes $Ti_3C_2Cl_2$ and $Ti_3C_2(OF)$ exhibit no discharge platforms, and only ~50% of capacities and energy densities of $Ti_3C_2Br_2$. These results emphasize the effectiveness of the Lewis-acidic-melt etching route for tuning the surface chemistry of MXenes, and also show promise for expanding the MXene family towards various applications.

**Introduction**

Two dimensional (2D) materials have attracted great interest due to their unique properties and wide range of applications. In 2011, a large family of 2D transition metal carbides and nitrides, the so-called MXenes, were discovered and expanded rapidly[1–4]. The various constituent elements and controllable surface chemistry provides room for a wide range of properties. MXenes exhibit extraordinary electronic, optical, and plasmonic properties, and show promise in applications such as supercapacitors, electromagnetic absorbing and shielding coating, catalysts, and carbon capture[5–12].

MXenes are predominantly produced by selective etching of the A layer atoms of MAX phases. The MAX phases are commonly described by the formula $M_{n+1}AX_n$ ($n =1-3$), where M is an early transition metal and X is carbon or nitrogen. A is an element traditionally from groups 13-16 but has recently been extended to transition elements[13,14]. The crystal structure of MAX phases can be viewed as stacking of edge-linked $M_6X$ octahedra (e.g. [$Ti_6C$]) layers with single atom layers of A-site elements (e.g., Al and Si). Owing to the relatively weak chemical bonding between the M and A elements, the A layer atoms can be removed from the MAX phases by chemical etchants, such as the widely used HF acid solution[1]. The resultant 2D nanostructured MXenes, have the formula of $M_{n+1}X_nT_x$, where $T_x$ is a surface terminal group originating from the etchant and solution, typically –OH, –O, and –F[1,4,15]. The composition and configuration of surface terminals influence the physical and chemical properties of MXenes, which have been well discussed by several theoretical works[12,16–18]. For instance, Xie *et. al.* pointed out that the Li-ion storage capacities of MXenes depend on the type of the surface terminals. Specifically, $Ti_3C_2$ MXene with –O terminal exhibits the highest theoretical Li-ion storage capacities[19]. However, tuning of surface terminal as well as exploring new surface terminals, although has been tried[11,20], remains a challenge experimentally. In 2019, a route for terminal tuning was investigated as MXenes with exclusive –Cl terminal ($Ti_3C_2Cl_2$ and $Ti_2CCl_2$), which were obtained through an approach of A-site-etching of MAX phases in Lewis acidic chloride melts[21,22]. Recently, we found that the Lewis-acidic-melt etching route is an general approach to produce MXenes with wide range of constituent[23]. This approach also implies the feasibility of tuning surface terminals of MXene and enriches the chemistry for enormous functionalities. In particular, if a greatly improved performance could be achieved through terminal tailoring, a wide range of new electrochemical applications would become achievable.

Here, $Ti_3C_2$ MXenes with unitary (–Cl, –Br and –I), binary (–ClBr, –ClI, and –BrI) and ternary (–ClBrI) halogen terminals were synthesized through the Lewis-acidic-melt etching route. These halogenated MXenes, as well as traditional MXene was produced by HF etching (denoted as $Ti_3C_2(OF)$), were assembled as cathodes for zinc ion batteries. Notably, unitary-halogen-terminated $Ti_3C_2Br_2$ and $Ti_3C_2I_2$ exhibit a discharge platform at approximately 1.6 V and 1.1 V with the specific capacities of 97.6 mAh·g$^{-1}$ and 135 mAh·g$^{-1}$. Multi-halogen-terminated $Ti_3C_2$(BrI) and $Ti_3C_2$(ClBrI) exhibit dual discharge platforms with the specific capacities of 117.2 mAh·g$^{-1}$ and 107.4 mAh·g$^{-1}$. In contrast, $Ti_3C_2Cl_2$ and $Ti_3C_2(OF)$ only show capacities of 46.5 and 51.7 mAh·g$^{-1}$, without platform. We further show that this exceptional electrochemical performance of the halogenated MXenes is attributed to the reversible redox behavior of –Br and –I terminals besides outmost Ti layers in $Ti_3C_2$. This work well reveals the role of active surface terminals in the electrochemical performance of MXenes, essential for tuning the properties and expanding the range of applications for MXenes.

**Results and discussion**

**Materials Characterization.** The XRD patterns in Fig. 1a shows the (0002) and (0004) peaks of as-obtained halogenated MXenes, as well as $Ti_3C_2(OF)$. Obvious difference in the angles of the (0004) peaks was observed due to the different *c* values of these MXenes. $Ti_3C_2Cl_2$ and $Ti_3C_2(OF)$ have intense (0002) peaks corresponding to the *c* values of respective 20.90 Å and 22.22 Å, in agreement with previous reports [1,21]. In contrast, the (0002) peaks of $Ti_3C_2Br_2$ and $Ti_3C_2I_2$ are less clear, but the (0004) peaks are strong enough to determine their respective *c* values of 23.33 Å and 25.00 Å, which are somewhat larger than the theoretical values calculated by DFT (23.02 Å and 24.24 Å). The increased interlayer spacing in $Ti_3C_2Br_2$ and $Ti_3C_2I_2$ can be explained by the increasing atom radius of the halogen group from fluorine to iodine. The *c* values of the binary- and ternary- halogen-terminated MXenes was in between that of the unitary-halogen-terminated MXenes, e.g. the $c$ values of $Ti_3C_2(BrI)$ (24.06 Å) is larger than that of $Ti_3C_2Br_2$ but smaller than $Ti_3C_2I_2$. The $c$ values determined form the (0004) peaks and elemental composition determined form EDS analysis (Fig S3-S9) of these halogenated MXenes were listed in Table 1. It is clear that tuning the terminals composition is an effective route to tune the interlayer spacing of MXenes. In addition, according to the full scale XRD patterns shown in Fig. S1, most of the non-basal-plane peaks of $Ti_3AlC_2$, e.g. the (104) peak at ~39°, were significantly diminished or even disappeared in these MXenes samples, indicating a complete removal of the single-atomic A layers.

$Ti_3C_2Br_2$ exhibit the typical accordion morphologies similar to previous as-etched $Ti_3C_2(OF)$ and $Ti_3C_2Cl_2$ MXenes (Fig. 1b), indicating that the van der Waals force between the single nanosheets of MXenes is still strong enough after etching-out of Al. Semi-quantitative analysis by EDS technique indicate atomic ratios of Ti:B ≈ 3:2 for $Ti_3C_2Br_2$ (Fig.S4), implicating nearly complete occupancy of halogen atoms at the surface terminal site. Fig. 1c shows a HAADF-STEM image and corresponding EDS elemental maps of $Ti_3C_2Br_2$. The Br element is uniformly distributed across the particle and no element segregation is observed. The HR-TEM image (Fig. 1d) shows that the $d$ value of the (0002) plane is ~11.7 Å, which agrees well with the XRD results. Fig. 1e and 1f is an atomically resolved STEM image with corresponding EDS elemental map revealing the Br terminals on the surface of $Ti_3C_2$. The brightness of atoms originates from mass-dependent scattering conditions and also demonstrates the successful functionalization by Br of the MXene surface. Fig. 1g, 1h and Fig. S10 shows a HAADF-STEM image and corresponding EDS elemental maps to show the morphology and element distribution of several other halogenated MXenes, i.e. $Ti_3C_2I_2$, $Ti_3C_2(ClBrI)$ and $Ti_3C_2(BrI)$. Note that for all MXenes, the halogen elements distributed uniformly, which attributes to satisfied accessibility of halogen anions in the molten salts.

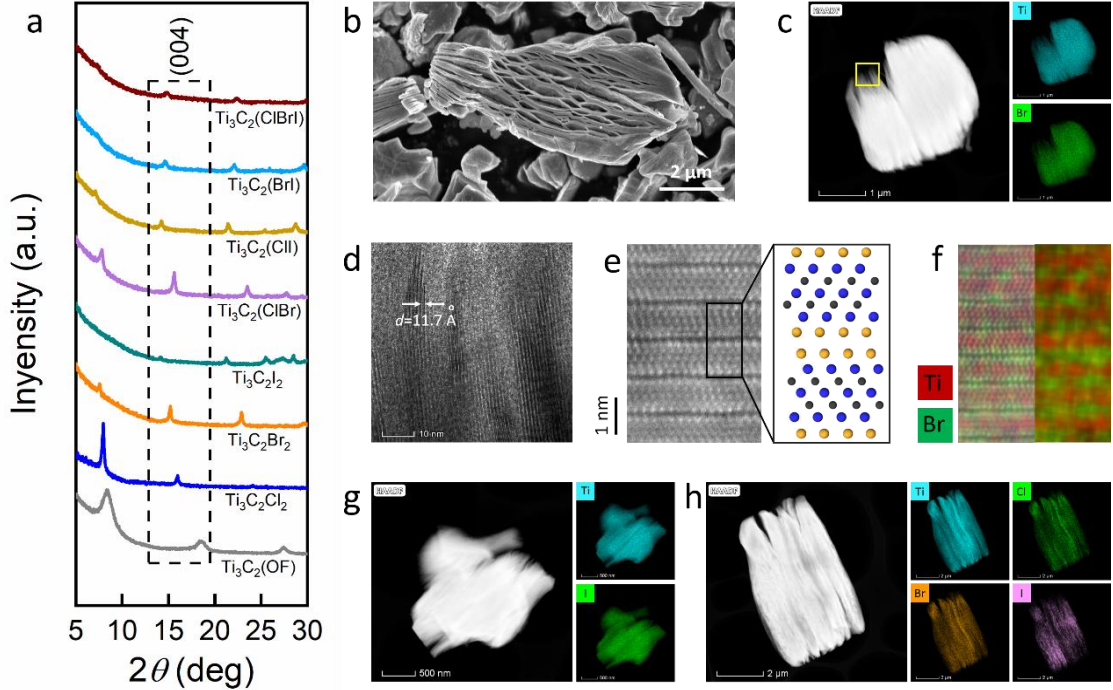

**Fig 1. Characterization of Ti$_3$C$_2$Br$_2$ and Ti$_3$C$_2$I$_2$.** (a) XRD patterns show the (0002), (0004) and (0006) peaks of the MXenes obtained in the present work. (b) SEM image of Ti$_3$C$_2$Br$_2$. (c) HAADF image and corresponding EDS map of Ti$_3$C$_2$Br$_2$. (d) HR-TEM image showing enlarged view of the marked area in Fig. 1d. (e, f) HR-STEM image and corresponding EDS map showing the atomic positions of Ti$_3$C$_2$Br$_2$. (g, h) HAADF image and corresponding EDS map showing the element distribution of Ti$_3$C$_2$I$_2$, and Ti$_3$C$_2$(ClBrI).

Table 1 Composition and *c* lattice parameter of the MXenes obtained in the present work

| Sample name | Composition | *c* lattice parameter (Å) |
|---|---|---|
| Ti$_3$C$_2$(OF) | – | 20.90 |
| Ti$_3$C$_2$Cl$_2$ | Ti/Cl=3/1.92 | 22.22 |
| Ti$_3$C$_2$Br$_2$ | Ti/Br=3/2.01 | 23.32 |
| Ti$_3$C$_2$I$_2$ | Ti/Br=3/1.89 | 25.00 |
| Ti$_3$C$_2$(ClBr) | Ti/Cl/Br=3/0.86/1.25 | 22.64 |
| Ti$_3$C$_2$(ClI) | Ti/Cl/I=3/0.31/1.55 | 24.87 |
| Ti$_3$C$_2$(BrI) | Ti/Br/I=3/0.99/1.02 | 24.06 |
| Ti$_3$C$_2$(ClBrI) | Ti/Cl/Br/I=3/0.58/0.76/0.68 | 23.50 |

Similar to our previous report on synthesizing Ti$_3$C$_2$Cl$_2$[21,23], the formation of Ti$_3$C$_2$Br$_2$ and Ti$_3$C$_2$I$_2$ can be described by the following simplified reaction:

$$\text{Ti}_3\text{AlC}_2 + 2.5\text{CuBr}_2 \rightarrow \text{Ti}_3\text{C}_2\text{Br}_2 + 2.5\text{Cu} + \text{AlBr}_3 \uparrow \quad (1)$$

$$\text{Ti}_3\text{AlC}_2 + 5\text{CuI} \rightarrow \text{Ti}_3\text{C}_2\text{I}_2 + 5\text{Cu} + \text{AlI}_3 \uparrow \quad (2)$$

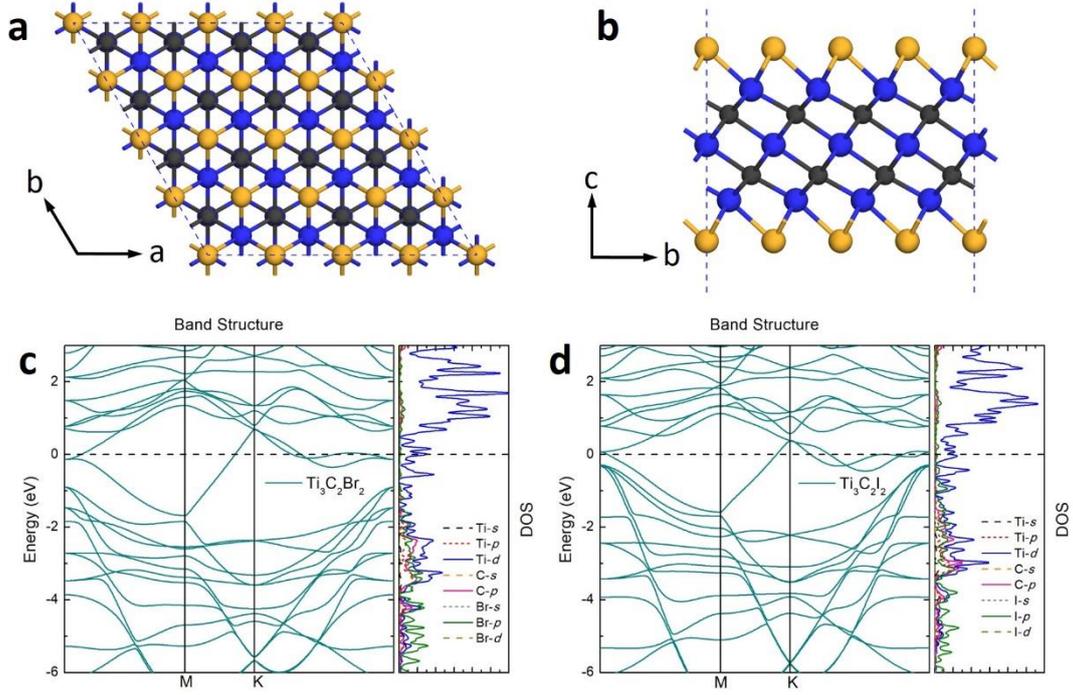

**Fig. 2 Configuration and electronic structure of $\text{Ti}_3\text{C}_2\text{Br}_2$ and $\text{Ti}_3\text{C}_2\text{I}_2$. (a, b)** Top view and side view of the monolayer $\text{Ti}_3\text{C}_2\text{Br}_2$ and $\text{Ti}_3\text{C}_2\text{I}_2$. **(c, d)** Electronic structure of $\text{Ti}_3\text{C}_2\text{Br}_2$ and $\text{Ti}_3\text{C}_2\text{I}_2$.

DFT calculations were carried out to investigate the atomic and electronic structure of two unitary-halogen-terminated MXenes: $\text{Ti}_3\text{C}_2\text{Br}_2$ and $\text{Ti}_3\text{C}_2\text{I}_2$. As shown in Fig. 2a and 2b, both the –Br and –I terminals preferentially occupy the fcc sites, which is similar to the Cl in $\text{Ti}_3\text{C}_2\text{Cl}_2$[22], as well as for O and F in $\text{Ti}_3\text{C}_2(\text{OF})$ [24,25]. The interplanar distance between the halogen terminals and the outmost Ti layer is 1.86 Å and 2.06 Å for $\text{Ti}_3\text{C}_2\text{Br}_2$ and $\text{Ti}_3\text{C}_2\text{I}_2$, which corresponds to a strong bonding of Ti–Br with bond length of 2.64 Å and a weak bonding of Ti–I with bond length of 2.83 Å. The bonding strength is further confirmed by the formation energy ($E_f$) of the halogen atoms on the MXene surface, which directly correlate with bonding strength between the terminal groups and the $\text{Ti}_3\text{C}_2$ matrix. The calculated $E_f$ of $\text{Ti}_3\text{C}_2\text{Br}_2$ and $\text{Ti}_3\text{C}_2\text{I}_2$ are respective −5.624 eV and −3.644 eV, which are less negative than the previously reported value of $\text{Ti}_3\text{C}_2\text{Cl}_2$ (−6.694 eV)[22], $\text{Ti}_3\text{C}_2\text{F}_2$ (−7.111 eV) and $\text{Ti}_3\text{C}_2\text{O}_2$ (−9.589 eV)[26]. The $E_f$ of the above MXenes follows the order of $\text{Ti}_3\text{C}_2\text{O}_2 < \text{Ti}_3\text{C}_2\text{F}_2 < \text{Ti}_3\text{C}_2\text{Cl}_2 < \text{Ti}_3\text{C}_2\text{Br}_2 < \text{Ti}_3\text{C}_2\text{I}_2$, indicating an opposite order of bonding strength between the outmost Ti atoms with the terminals. The formation energy of the binary- and ternary-halogen-terminated MXenes is yet need further investigate, while their value are expected to be superposition of the $E_f$ of unitary-halogen-terminated l MXenes in certain model as the alloy materials[27 28].

Fig. 2c and 2d show the electronic structure of $Ti_3C_2Br_2$ and $Ti_3C_2I_2$, including the band structure and projected density of states (DOS). Both $Ti_3C_2Br_2$ and $Ti_3C_2I_2$ are metallic in nature with a finite density of states at the Fermi level predominately contributed by Ti *d*-orbitals. In addition, the calculations show a strong hybridization between Ti *d*-, C *p*- and Br/I *p*-orbitals in the energy range of -2 to -6 eV, indicating the strong coordination of outmost Ti atoms with carbon and halogen atoms in the form of edge-sharing octahedra [$TiC_3T_3$] (T=Br and I) in these halogenated MXenes.. The hybridization near the surface of the halogenated MXenes provide the possibility to enrich their surface chemistry. Moreover, the highly polar bonds near surface also induce van der Waals interaction between the MXene layers, which explains the fact that the all these halogenated MXenes display accordion morphology.

**Electrochemical Performance**. The dependence of electrochemical behavior on the terminal compositions are shown in the cyclic voltammetry (CV) curves of halogenated MXenes including $Ti_3C_2(OF)$, $Ti_3C_2Cl_2$, $Ti_3C_2Br_2$, $Ti_3C_2I_2$, $Ti_3C_2(BrI)$ and $Ti_3C_2(ClBrI)$ (Fig. 3a-3f). No redox peaks are observed in CV curves of $Ti_3C_2(OF)$ and $Ti_3C_2Cl_2$, indicating no conversion reaction occurs. This fact is in agreement with the previous reported pseudo-capacitive behaviors of MXenes, which are usually accompanied by a pair of broad redox peaks due to rapid ions insertion/extraction, whereas no obvious redox reaction involve the constituent of MXenes[5,8,29,30]. In contrast, characteristic redox peaks was observed in CV curves of both $Ti_3C_2Br_2$ and $Ti_3C_2I_2$ MXenes: 1.55/1.65 V peaks for $Ti_3C_2Br_2$ and 1.05/1.15 V peaks for $Ti_3C_2I_2$ (*vs* $Zn/Zn^{2+}$) are assigned to the revisable conversion of $Br^-/Br^0$ and $I^-/I^0$, respectively. The reaction is rationally proposed as below:

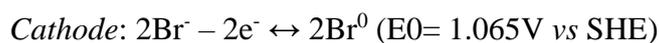
*Cathode*: $2Br^- - 2e^- \leftrightarrow 2Br^0$ (E0= 1.065V *vs* SHE)

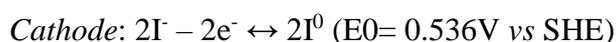
*Cathode*: $2I^- - 2e^- \leftrightarrow 2I^0$ (E0= 0.536V *vs* SHE)

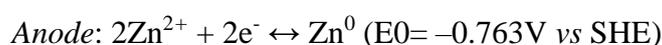
*Anode*: $2Zn^{2+} + 2e^- \leftrightarrow Zn^0$ (E0= –0.763V *vs* SHE)

For the multi-halogen-terminated MXenes, i.e. $Ti_3C_2(BrI)$ and $Ti_3C_2(ClBrI)$, dual redox peaks was distinguished in the CV curves. For an instance, the 1.33/1.45 V peaks and 1.05/1.15 V peaks are respectively assigned to the redox reaction of $Br^-/Br^0$ and $I^-/I^0$ in MXene $Ti_3C_2(BrI)$. The CV curves of $Ti_3C_2(ClBrI)$ is almost the same as that of $Ti_3C_2(BrI)$, which corroborates that the Cl terminals contribute no electrochemical activity in the present system. The reduced redox potential of $Br^-/Br^0$ in multi-halogen-terminated halogenated MXenes is reasonable in a complex coordinating octahedra $[TiC_3Br_nI_{3-n}]$ (n=1 to 2). Since radius of $I^-$ ($r_{I^-}$=0.220 nm) is larger than that of $Br^-$ ($r_{Br^-}$=0.196 nm), the $[TiC_3Br_nI_{3-n}]$ octahedra structure must be distorted to accommodate these two kinds of halogen atoms. Consequently, Br atoms are pushed away by surrounding I atoms from outmost Ti atoms according to hard-ball model in crystallography, and become more active to lose and obtain electrons during redox reaction.

The galvanostatic charge-discharge profiles (GCD) in Fig. 3g also confirms the different electrochemical behavior of the MXenes. The GCD curves of $Ti_3C_2Cl_2$ and $Ti_3C_2(OF)$ show similar features, i.e. close to two straight lines without any charge-discharge platform. The corresponding discharge capacities are only 46.5 and 51.7 mAh·g$^{-1}$ at 0.5 A·g$^{-1}$. The capacities of $Ti_3C_2Cl_2$ and $Ti_3C_2(OF)$ are contributed from the pseudo-capacitive of MXene as confirmed in previous reports[31–33]. In contrast, $Ti_3C_2Br_2$ and $Ti_3C_2I_2$ show obvious and flat discharge platforms at ~1.6 V and ~1.1 V with the capacities of 97.6 and 135 mAh·g$^{-1}$, respectively. In agreement with the CV curves, both $Ti_3C_2(BrI)$ and $Ti_3C_2(ClBrI)$ show dual flat discharge platforms at ~1.4 V and ~1.1 V. $Ti_3C_2(BrI)$ show capacity of 117.2 mAh·g$^{-1}$, while $Ti_3C_2(ClBrI)$ show lower capacity of 106. 7 mAh·g$^{-1}$ due to the effect of –Cl.

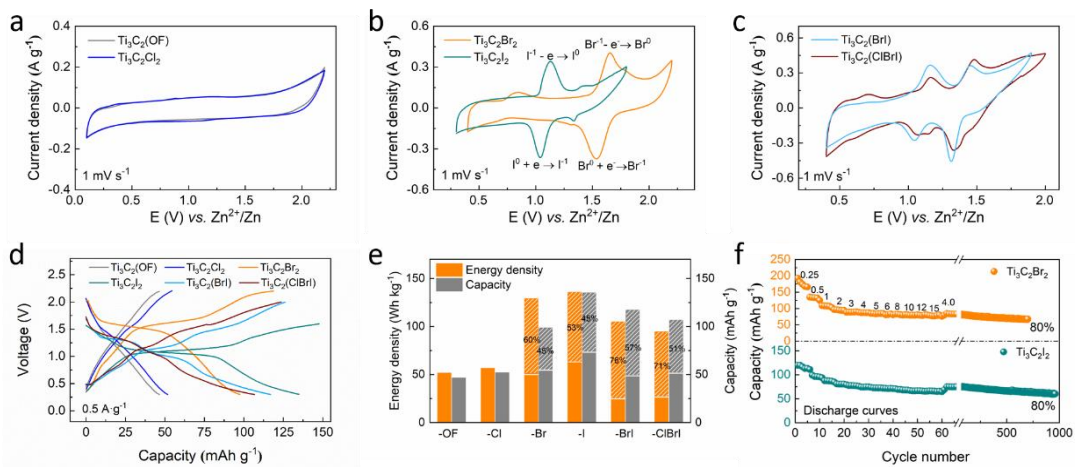

**Fig. 3 Electrochemical performance of halogenated MXenes. (a-c)** Cyclic voltammogram of $Ti_3C_2(OF)$, $Ti_3C_2Cl_2$, $Ti_3C_2Br_2$, $Ti_3C_2I_2$, $Ti_3C_2(BrI)$, and $Ti_3C_2(ClBrI)$ at a scan rate of 1 mV s$^{-1}$. **(d)** Typical GCD curves of $Ti_3C_2$ MXenes with different terminals at the current density of 0.5 A g$^{-1}$. **(e)** The energy density and capacity of $Ti_3C_2$ MXenes with different terminals. The dash area presents the contribution from discharge platform region. **(f)** Rate capability and long-term cycling performance of $Ti_3C_2Br_2$ and $Ti_3C_2I_2$.

As shown in Fig. 3h, the total capacities of $Ti_3C_2Br_2$ and $Ti_3C_2I_2$ are composed of two parts of contributions. One part is attributed to the pseudo-capacitive that has similar value as that of $Ti_3C_2Cl_2$ and $Ti_3C_2(OF)$. The other part, ~45% of the whole capacities, are contributed from the discharge platform. Furthermore, $Ti_3C_2Br_2$ and $Ti_3C_2I_2$ show a high energy density of 129.0 and 135.6 Wh·Kg$^{-1}$, which are almost ~200% of that of $Ti_3C_2Cl_2$ and $Ti_3C_2(OF)$. Note that all the extra energy density of $Ti_3C_2Br_2$ and $Ti_3C_2I_2$ originates from the high voltage platform region contribution (Fig. 3f), which satisfies the current expectation for aqueous batteries. $Ti_3C_2(BrI)$ and $Ti_3C_2(ClBrI)$ shown energy density of 104.85 and 94.6 Wh·Kg$^{-1}$. The energy density of $Ti_3C_2(BrI)$ and $Ti_3C_2(ClBrI)$ is lower than that of $Ti_3C_2Br_2$ and $Ti_3C_2I_2$, while the energy density originates from the platform region is much higher due to the dual platforms. The above results demonstrate that that the electrochemical performance of the multi-halogen terminated MXenes is the superposition of the unitary terminals MXenes in certain degree, thus the surface terminal tailing provide an effective strategy to tune the electrochemical performance of MXenes.

$Ti_3C_2Br_2$ and $Ti_3C_2I_2$ also exhibit excellent rate performance (Fig. 3f and Fig. S15). The respective discharge capacities of $Ti_3C_2Br_2$ and $Ti_3C_2I_2$ are 181 and 120 mAh·g$^{-1}$ at 0.25 A·g$^{-1}$. Even under the ultrahigh current density of 15 A·g$^{-1}$, 45% and 56% of capacity retentions are retained for $Ti_3C_2Br_2$ and $Ti_3C_2I_2$. Meanwhile, voltage platform preserved as well as the high energy density. With the capacity decay of 20% as a simple evaluation of reliability, $Ti_3C_2Br_2$ and $Ti_3C_2I_2$ can stabilize the cycling up to 700 and 1000 rounds, which implicates decent lifetime of these Zinc-ion battery with halogenated MXene cathodes.

We have summarized the discharge voltage platform of various reported aqueous MXene-based energy storage systems including $K^+$, $H^+$, $Na^+$, $NH_4^+$, $Li^+$, $Mg^{2+}$, $Zn^{2+}$, $Al^{3+}$, and some widely studied cathodes in aqueous Zn batteries including Mn-, V-, $MoS_2$, and polymer (Table S1 and Fig. S18). The current halogenated MXene cathodes provide new solution to following two challenges: i) Most of the cathodes used in aqueous Zn batteries can not realize a low charging-discharging platform. ii) In the aqueous environment, previously reported MXene electrodes only exhibit low pseudo-capacity without contribution from terminal groups.

**Energy Conversion Mechanism.** In order to demonstrate the energy conversion mechanism of the halogenated MXenes, we choose unitary-halogen-terminated MXenes, $Ti_3C_2Br_2$ and $Ti_3C_2I_2$, to investigate their structure and composition change at different charge/discharge state. As shown in Fig. 4a, upon charging, a broad Raman shift at 260–270 cm$^{-1}$ corresponding to $Br_n^-$ ($3 \leq n \leq 5$) was detected in the $Ti_3C_2Br_2$ electrode[34–36]. Considering the reversible reaction, $Br^- + (n-1)Br^0 \leftrightarrow Br_n^-$, the formation of $Br_n^-$ indicates the generation of $Br^0$ ($Br_2$) due to the oxidation of $Br^-$. Upon discharging, the peaks of $Br_n^-$ disappeared due to the recovery of $Br_2$ into $Br^-$ after capture electrons. As regard to $Ti_3C_2I_2$ electrode, upon charging, a weak Raman shift at 114 cm$^{-1}$ was observed which corresponds to the formation of $I_3^-$ (Fig. 4d)[37,38], indicating the generation of $I^0$ ($I_2$) during the oxidation of $I^-$. Upon discharging, the Raman shift at 114 c m$^{-1}$ disappears gradually, indicating the recovery of $I^-$ ion from $I^0$, which is similar to that of $Ti_3C_2Br_2$. The valence change of Br and I element was further confirmed by XPS analysis. As shown in Fig. 4b, the Br 3d peak shifts from 69.2 eV to 70.5 eV during the change process, indicating an increased valence of Br element caused by losing electron. In the discharge process, the Br 3d peak shifts in the opposite direction, indicating a decreased valence of Br element caused by obtaining electron. A similar phenomenon was observed in the I 3d XPS analysis of $Ti_3C_2I_2$ electrode (Fig. 4f). The I $3d_{5/2}$ peak increases from 619. 5eV to 620.2 eV during the change process and decreased gradually during the discharge process, which confirms the oxidation and reduction of I element.

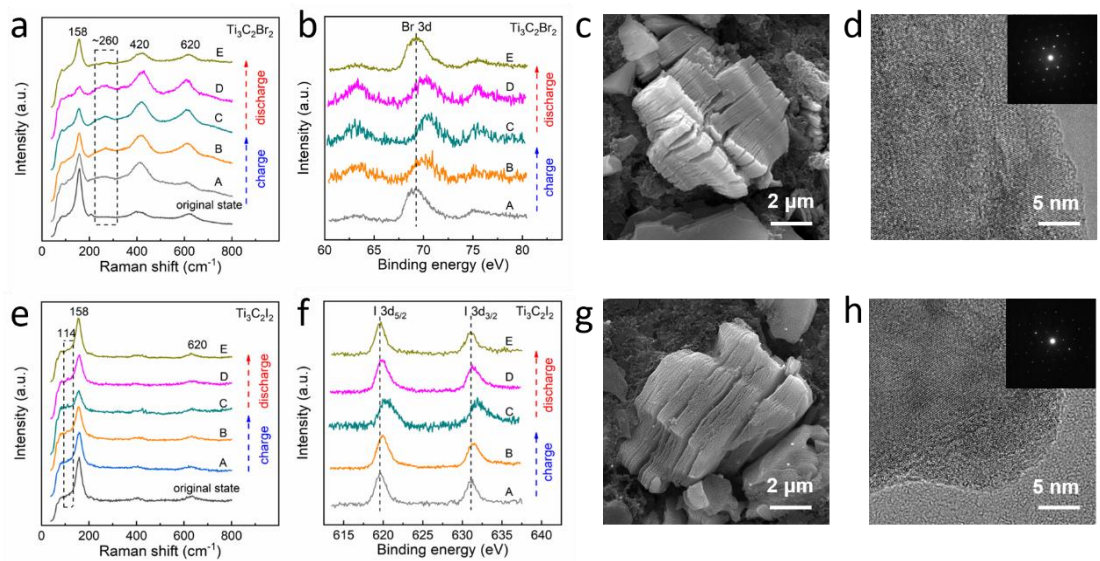

**Fig. 4 Structure and composition change of the $Ti_3C_2Br_2$ and $Ti_3C_2I_2$ electrode after cycles.** (a, d) *Ex-situ* Raman spectra of $Ti_3C_2Br_2$ and $Ti_3C_2I_2$ electrode at charge/discharge state. For $Ti_3C_2Br_2$, the voltage of each point is A: 0.3 V, B: 1.0 V, C: 1.6 V, D: 0.9 V, E: 0.3 V. For $Ti_3C_2I_2$, the voltage of each point is A: 0.3 V, B: 1.3 V, C: 2.2 V, D: 1.2 V, E: 0.3 V. (b, f) Br 3d and I 3d XPS analysis of $Ti_3C_2Br_2$ and $Ti_3C_2I_2$ electrode at charge/discharge state. (c, g) SEM images of $Ti_3C_2Br_2$ and $Ti_3C_2I_2$ after cycles. (d, h) HR-TEM image of $Ti_3C_2Br_2$ and $Ti_3C_2I_2$ after cycles, the insertion is the corresponding SAED pattern.

The structure and composition of $Ti_3C_2Br_2$ and $Ti_3C_2I_2$ electrodes after 700 and 1000 cycles were further investigated by SEM and TEM. As shown in Fig. 4b and 4e, both the $Ti_3C_2Br_2$ and $Ti_3C_2I_2$ electrode retain the accordion morphology after cycles. Both the $Ti_3C_2Br_2$ and $Ti_3C_2I_2$ remains crystalline according to the HR-TEM image and the corresponding SAED pattern (Fig. 4c and 4f), indicating their affordable structure stability. The EDS analysis indicates that the main elemental composition of the $Ti_3C_2Br_2$ after cycles is Ti/C/O=37.2: 29.5: 28.8 in atomic ratio, whereas the Br ratio decreased dramatically to 1.4 at% (Fig. S11). According to the HAADF image and corresponding EDS map of $Ti_3C_2Br_2$ sample (Fig. S12), all elements distribute uniformly and no segregation occurs. As regard to $Ti_3C_2I_2$, the main elemental composition of the sample is Ti/C/O=29.5: 23.5: 35.4 in atomic ratio, and the content of I decreased to 5.3 at% (Fig. S13). Notably, segregation iodine element is observed in the cycled $Ti_3C_2I_2$ electrode (Fig. S14), which accounts for their solid state and resultant low diffusion rate of iodine as compared with liquid-state bromine species.

Base on above analysis, the energy conversion process of the –Br and –I terminated MXenes electrodes take advantage of their redox reaction of halogen groups confined in $Ti_3C_2$ layers. Upon charging, halogen ions in a negative covalent state lose electrons and are oxidized into near-zero-valent state. At the same time, the $Zn^{2+}$ ions in the electrolyte obtain electrons and deposited on the anode surface. Upon discharging, the above reactions are reversed. In contrast, the –O, –F, and –Cl groups in $Ti_3C_2Cl_2$ and $Ti_3C_2(OF)$ MXenes upon charging at much high voltage should transform into gases, e. g. $O_2$, $F_2$, and $Cl_2$, which would evaporate out of the interlayers of $Ti_3C_2$ layer as soon as form. Therefore, even the redox potentials are satisfied for the –O, –F, and –Cl terminals (O: 1.229 V *vs* SHE, F: 2.866 V *vs* SHE, Cl: 1.358 V *vs* SHE), no revisable conversion reaction and corresponding platform could be realized.

**Conclusion.**

Novel halogenated $Ti_3C_2T_2$ (T= –Br and –I, or their combination) MXenes were synthesized through a Lewis-acidic-melt etching route. These halogenated MXenes show extraordinary performance as cathodes for aqueous zinc ion batteries with merits of a stable flat discharge platform, considerable capacity, high energy density and good rate performance, which makes them highly competitive among most of the cathodes used in aqueous zinc ion batteries. In contrast, the previously discovered $Ti_3C_2T_x$ MXenes (T= –Cl, –F, –O and –OH) showed far less capacities with no discharge platform. This work demonstrated the large room of surface chemistry tuning for the improvement of electrochemical performance of MXenes, as well as other functional applications.

**Methods**

**Materials preparation.** $Ti_3AlC_2$ powder (particle size ~20 μm) produced by a molten-salt method was used as precursor to synthesize the $Ti_3C_2T_x$ MXenes in the present work. The $Ti_3C_2$ MXenes with halogen terminals were prepared by a reaction between $Ti_3AlC_2$ with the corresponding copper halide molten salts. The composition of the starting materials was listed in Table 2. The starting materials were mixed thoroughly using a mortar under the protection of nitrogen in a glovebox. Then the as-obtained mixture powder was taken out from the glove box and placed into an alumina crucible. The alumina crucible was loaded into a tube furnace and heat-treated at 700 °C for 7 h under the protection of argon gas to accomplish the reaction. Note that the reaction product is a mixture of the MXenes, residual salt and by-product copper. The products were washed by deionized water to remove the residual $CuCl_2$ and $CuBr_2$, and further washed by a $NH_4Cl/NH_3·H_2O$ solution to remove the copper and CuI. The comprehensive characterization (XRD in Fig. S2, SEM and EDS in Fig. S3-S9) of the product before and after $NH_4Cl/NH_3·H_2O$ washing was shown supporting information. Note that the copper-removing process does not have much effect on the composition and morphology of MXenes. After the $NH_4Cl/NH_3·H_2O$ solution washing, the products were dried at 40°C in vacuum to obtain the $Ti_3C_2$ MXenes with halogen terminals. In addition, the traditional $Ti_3C_2$ MXene with a mixture terminal of –F, –O, and –OH( denoted as $Ti_3C_2(OF)$), was produced by a typical HF etching process [1].

Table 2. The composition of the starting materials for preparing the halogenated MXenes.

| Sample name | Composition of starting materials (molar ratio) |
|---|---|
| $Ti_3C_2Cl_2$ | $Ti_3AlC_2/CuCl_2$=1/4 |
| $Ti_3C_2Br_2$ | $Ti_3AlC_2/CuBr_2$=1/4 |
| $Ti_3C_2I_2$ | $Ti_3AlC_2/CuI$=1/6 |
| $Ti_3C_2(ClBr)$ | $Ti_3AlC_2/CuCl_2/CuBr_2$=1/1/4 |
| $Ti_3C_2(ClI)$ | $Ti_3AlC_2/CuCl_2/CuI$=1/1/6 |
| $Ti_3C_2(BrI)$ | $Ti_3AlC_2/CuBr_2/CuI$=1/1/6 |
| $Ti_3C_2(ClBrI)$ | $Ti_3AlC_2/CuCl_2//CuBr_2/CuI$=1/1/1/6 |

**Characterization.** Scanning electron microscopy (SEM) characterization was performed in a thermal field emission scanning electron microscope (Thermo scientific, Verios G4 UC) equipped with an energy-dispersive spectroscopy (EDS) system. Transmission electron microscopy (TEM) and scanning transmission electron microscopy (STEM) were performed in the Linköping mono-chromated, high-brightness, double-corrected FEI Titan[3] 60–300 operated at 300 kV, equipped with a SuperX EDS system. X-ray diffraction (XRD) analysis of the powders was performed using a Bruker D8 ADVANCE X-ray diffractometer with Cu Kα radiation at a scan rate of 2 °/min. Raman spectra measured by a Ramoscope (Renishaw inVia Reflex) with the excitation wavelength of 532 nm. X-ray photoelectron spectra (XPS) were performed in an XPS system (Axis Ultra DLD, Kratos, UK) with a monochromatic Al X-ray source. The binding energy (BE) scale were assigned by adjusting the C 1s peak at 284.8 eV.

**Electrochemical Measurements.** The cathode was prepared by mixing MXene, black carbon and PVDF (poly(vinylidene difluoride)) with a mass ratio of 7:2:1. Carbon cloth was selected as the current collector. Fresh Zn foil with the 50 μm thickness was directly employed as the anode, which conducts the deposition-dissolution process with a demarcation voltage point at -0.73 V *vs* Ag/AgCl (SI, Fig. S16). The aqueous electrolyte was obtained by dissolving 0.21 mol LiTFSI and 0.07 mol LiOTf and 0.01 mol $Zn(OTf)_2$ salt into 10 ml deionized water, followed by sealing and vigorous stirring for 24 h at 45 °C. Highly concentrated $Li^+$ ions bond most of the water in the electrolyte, which can widen the decomposition voltage window of electrolyte to more than 3.5 V, ranging from -1.5 to 2.0 V *vs* Ag/AgCl (SI, Fig. S17). CV test of Zn anode was carried out by CHI 660 electrochemical workstation based on three electrodes method, where Ag/AgCl acts as the reference electrode and two Pt (1cm×1cm) foils act as the counter electrode and work electrode. For the linear sweep voltammetry (LSV) test of electrolyte, the method is the same as above, two Pt foils (1cm×1cm) and Ag/AgCl were employed as working, counter and reference electrodes. At this time, in order to avoid the influence of Zn ions deposition/dissolution on the results, the electrolyte used here does not contain $Zn(OTf)_2$ salt. The coin-type (CR2032) batteries were assembled for all electrochemical measurements, using commercial non-woven fibrous membrane as the separator. Electrochemical profiles of the full battery were performed using the Land 2001A battery testing system and CHI 660 electrochemical workstation.

**Computational Methods.** First-principles density-functional theory (DFT) calculations were performed in the CASTEP module [39,40]. The energy cutoffs were set as 550 eV to describe the electronic wave functions based on the projected-augmented wave (PAW) approach, and the generalized gradient approximation (GGA) as implemented by Perdew−Burke−Ernzerhof (PBE) was used as exchange-correlation functional[41–44]. For the structural optimization, the Brillouin zone (BZ) is sampled using a set of Γ-centered $12 \times 12 \times 1$ k-points, the total energy change is smaller than $1\times10^{-7}$ eV/atom, and the maximum force on each atom is less than $1\times10^{-4}$ eV/Å.

The formation energy ($E_f$) of the halogen-terminated MXenes was calculated as

the following formula:

$$E_f = E(Ti_3C_2T_2) - E(Ti_3C_2) - E(T_2)$$

where $E(Ti_3C_2T_2)$ is the total energy of the $Ti_3C_2T_2$ MXene, $E(Ti_3C_2)$ is the energy of the non-terminated surface, and $E(T_2)$ is the energy of a free molecule of the terminals ($Br_2$ and $I_2$).

**Acknowledgements**

This study was supported financially by the National Natural Science Foundation of China (Grant No. 21671195 and 51902320), and China Postdoctoral Science Foundation (Grant No. 2018M642498). QH thanks International Partnership Program of Chinese Academy of Sciences (Grant No. 174433KYSB20190019), Leading Innovative and Entrepreneur Team Introduction Program of Zhejiang (Grant No. 2019R01003), and Ningbo top-talent team program for financial support. This study was also supported by GRF under Project N_CityU11305218 and the Science Technology and Innovation Committee of Shenzhen Municipality (the Grant No. JCYJ20170818103435068). The Knut and Alice Wallenberg Foundation is acknowledged for support of the electron microscopy laboratory in Linköping, a Fellowship grant (P.E.) and a project grant (KAW 2015.0043). P. O. Å. P. and J. L. acknowledge the Swedish Foundation for Strategic Research (SSF) for project funding (EM16-0004) and the Research Infrastructure Fellow (RIF 14-0074). We also acknowledge the Swedish Government Strategic Research Area in Materials Science on Functional Materials at Linköping University (Faculty Grant SFO-Mat-LiU No. 2009 00971).